\documentclass[aps,PRL,twocolumn,groupedaddress,floats]{revtex4}
\usepackage{amssymb}
\usepackage{graphicx}
\usepackage[dvipdfm]{hyperref}

\begin{document}

\title{Electronic structure reconstruction: the driving force behind the magnetic and structural transitions in NaFeAs}

\author{C. He$^{1}$, Y. Zhang$^{1}$,  B. P. Xie$^{1}$,  X. F. Wang$^{2}$,  L. X. Yang$^{1}$,  B. Zhou $^{1}$, F. Chen$^{1}$, M. Arita$^{3}$, K. Shimada$^{3}$, H. Namatame$^{3}$, M.
Taniguchi$^{3}$,  X. H. Chen$^{2}$,  J. P. Hu$^{4}$}
\email{hu4@physics.purdue.edu}
\author{D. L. Feng$^{1}$}
\email{dlfeng@fudan.edu.cn}

\affiliation{$^1$Department of Physics, Surface Physics Laboratory (National Key Laboratory), and Advanced Materials Laboratory, Fudan University, Shanghai 200433, People's Republic of China}

\affiliation{$^2$ Department of Physics,  University of science and technology of China, Hefei, Anhui 230027, People's Republic of China}

\affiliation{$^3$Hiroshima Synchrotron Radiation Center and Graduate
School of Science, Hiroshima University, Hiroshima 739-8526, Japan.}

\affiliation{$^4$ Department of Physics, Purdue University, West
Lafayette, Indiana 47907, USA}

\date{\today}

\begin{abstract}

The electronic structure of NaFeAs is studied with  angle resolved photoemission
spectroscopy on high quality
single crystals. Large portions of the band structure start to shift around the structural transition temperature, and smoothly evolve  as the temperature lowers through the spin density wave transition. Moreover, band folding due to magnetic order emerges around structural transition. Our observation provides direct evidence that the  structural and magnetic transitions  share the same origin, and are both
driven by the electronic structure reconstruction in Fe-based
superconductors, instead of Fermi surface nesting.
\end{abstract}


\maketitle


Iron based high temperature superconductors, like the cuprates, are at the vicinity of an ordered magnetic phase \cite{Hosono,XHChen,ZXZhao}, and magnetism has been suggested to be crucial for the superconductivity in both cases. A unique feature  associated with the magnetic  transition in iron pnictides/chalcogenides is that the magnetic or spin density wave (SDW) transition is always accompanied by a structural transition.   Although theoretically, it has been suggested that  the  structural transition may be driven magnetically \cite{ChenFang}, it does occur at higher temperature in many cases, and even isotope effects have been reported for the magnetic/structural transitions  \cite{XianhuiNature}. So far, there is no hard experimental evidence to establish the relationship between the structural transition and  the magnetic transition, but their ubiquitous co-occurrence  makes it crucial to understand their nature and origin, in order to unveil the mystery of iron-based superconductors.

For BaFe$_2$As$_2$ and other parent compounds of the so called ``122" series of iron pnictides and Fe$_{1+y}$Te, the structural transition occurs simultaneously with the magnetic transition \cite{QHuang,JunZhao,WeiBao,ShiliangLi2}. It is thus hard to separate the effects of lattice and magnetic transitions there. For LaOFeAs (a representing parental compound of the ``1111" series) \cite{PCDai} and NaFeAs (representing the ``111" series) \cite{ShiliangLi}, the structural transition precedes the SDW transition. However, the charge redistribution on the surface of the ``1111" series hampers direct measurements of its bulk electronic structure \cite{DHLu}. Without these constraints, NaFeAs provides an ideal opportunity to address what happens to the electronic structure between the lattice and SDW transitions that has not been accessed  before.

In this Letter, we report angle resolved photoemission spectroscopy (ARPES)  measurements of the detailed electronic structure of NaFeAs. We show that there is strong band renormalization, and the SDW is not caused by Fermi surface nesting, and rather the electronic energy is saved through band reconstruction. The entire band structure including bands well below Fermi surface participates in the reconstruction process.  Such a transition of the electronic structure  emerges at the structural transition temperature, and it smoothly evolves through the SDW transition temperature. Moreover, band folding is observed even at the structural transition temperature, which suggests   short-ranged magnetic ordering has taken place at this temperature. These results establish that both the lattice structural  transition and the magnetic transition are driven by the electronic structure reconstructions in Fe-based superconductors.


High quality NaFeAs single crystals were synthesized by a NaAs flux method.  Na pieces and As powder was mixed and heated to 250$^{\circ}$C for 3 hours. The Fe and NaAs powders were loaded into an alumina crucible and then sealed in a Ta crucible.  They were heated to 900$^{\circ}$C, then slowly cooled to 600$^{\circ}$C, and to room temperature after shutting off the power of the furnace. Resistivity measurements confirm that there is a superconducting transition at 8K,  a SDW transition at $T_{N}$=39K, and a structural transition at $T_{S}$=54K \cite{GFChen,ShiliangLi}. ARPES data were taken with circularly polarized 19\,eV photons at the Beamline 9 of Hiroshima synchrotron radiation center (HSRC) with a Scienta R4000 electron analyzer; the energy resolution was 9\,meV, and the angular resolution was 0.3 degree.  Sample was measured within two hours after it was cleaved \textit{in situ} under ultra-high-vacuum of $3\times10^{-11}$\textit{torr}. Aging effects are strictly monitored, so that data reflect the intrinsic electronic structure.


\begin{figure}[t!]
\includegraphics[width=8cm]{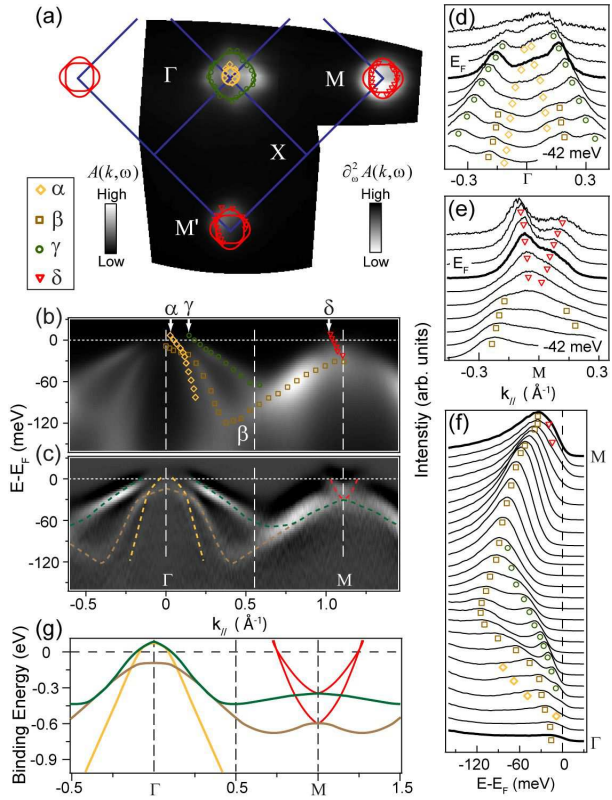}
\caption{(Color online) Electronic structure of NaFeAs at 60K. (a) Photoemission intensity map at the Fermi energy ($E_F$) integrated over [$E_F$-5~meV,$E_F$+5~meV], the marks are measured Fermi crossings, while the curves are the fitted Fermi surfaces. (b) Photoemission intensity $A(k,\omega)$ along the $\Gamma-M$ direction. (c) The second derivative of photoemission intensity with respect to energy [$\partial_{\omega}^{2}A(k,\omega)$] for data in panel b. The dashed lines are local minimum locus to indicate the band position. For simplicity, the bands are denoted ignoring the band crossings, and each band is represented by the same color hereafter when necessary. (d) and (e) Momentum distribution curves (MDCs) around $\Gamma$ and $M$ respectively; each MDC has been individually normalized by its integrated weight to highlight weak features. (f) Selected energy distribution curves (EDCs) along the $\Gamma-M$ direction. (g) Calculated band structure reproduced from Ref. \cite{theory}. } \label{normal}
\end{figure}

The normal state photoemission intensity map at the Fermi energy is shown in Fig.~\ref{normal}a. Two holelike Fermi pockets near $\Gamma$ and two electron-like Fermi surfaces near $M$ are observed. The photoemission intensity along the $\Gamma-M$ direction and its second derivative with respect to energy are shown in Figs.~\ref{normal}b and ~\ref{normal}c respectively. In Fig.~\ref{normal}b, four bands ($\alpha$, $\beta$, $\gamma$, and $\delta$) are determined by tracking the peaks in momentum and energy distribution curves (MDCs and EDCs) in Figs.~\ref{normal}d-f, and they agree with the band structure determined by the local minimum locus in Fig.~\ref{normal}c. The overall normal state electronic structure of NaFeAs is similar to other pnictides. However, a distinct feature for NaFeAs is that three bands near $\Gamma$ are clearly observed here, while there are only two features observed in the paramagnetic state of systems like BaCo$_x$Fe$_{2-x}$As$_2$   or Ba$_{1-x}$K$_x$Fe$_{2}$As$_2$ \cite{HongDing1,HongDing2}.
Fig.~\ref{normal}g reproduces the band structure calculation of NaFeAs in Ref. \cite{theory}, which qualitatively agrees with the measured band structure. However quantitatively, the experimental bandwidth for $\beta$ and $\gamma$  is renormalized by a factor of about 5.4 and 6.5 respectively from the calculated ones, while this factor is $3\sim4$ for other iron pnictides and iron chalcogenides  \cite{chenf}, indicative of strong correlations in NaFeAs. Moreover, only one of the two sets of electron-like bands is observed near $M$ or $M'$, due to the matrix element effects.

\begin{figure}[t]
\includegraphics[width=8cm]{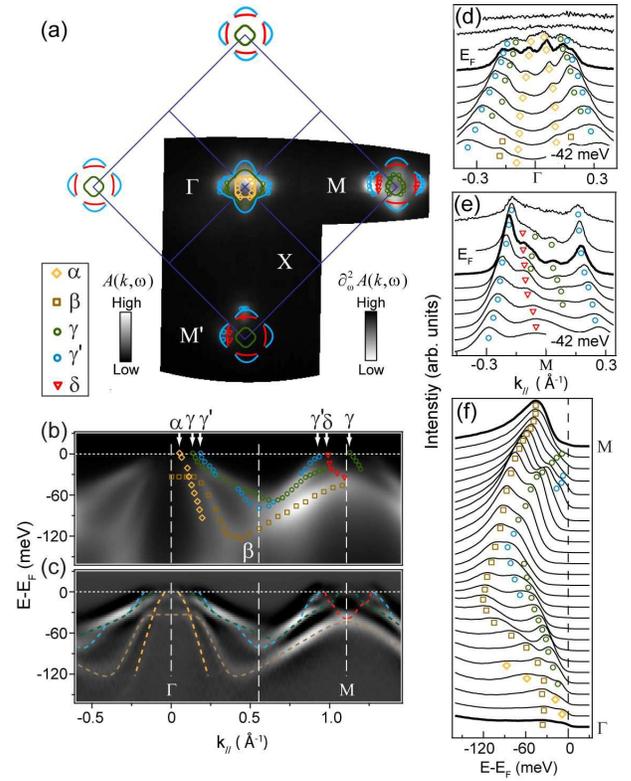}
\caption{(Color online)  SDW state electronic structure  of NaFeAs. (a) Photoemission intensity map at the Fermi energy integrated over [$E_F$-5~meV,$E_F$+5~meV], the marks are measured Fermi crossings, while the curves are the fitted Fermi surfaces. (b) Photoemission intensity along the $\Gamma-M$ direction. (c) The second derivative of data in panel b with respect to energy. The dashed lines are local minimum locus to indicate the band position. (d) and (e)  MDCs of data around $\Gamma$ and $M$ respectively, and each MDC has been individually normalized by its integrated weight. (f) Selected  EDCs along the $\Gamma-M$ direction. Data were taken at 10K.
}\label{SDW}
\end{figure}

\begin{figure}[t!]
\includegraphics[width=7cm]{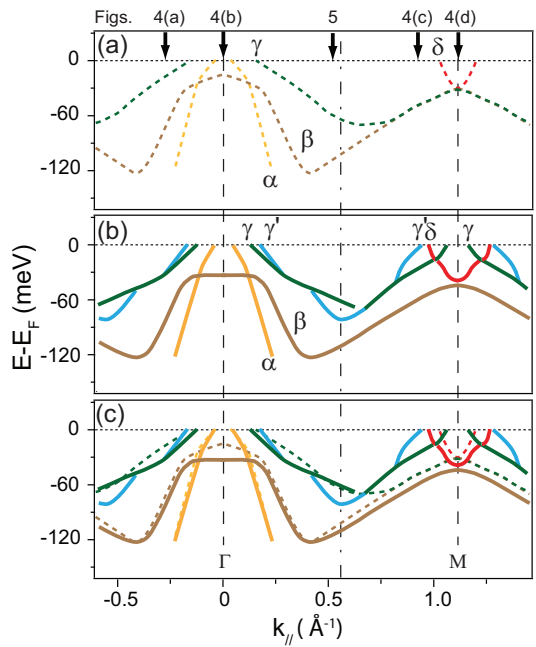}
\caption{(Color online) Band structure of NaFeAs determined experimentally at (a) the normal state, and (b) the SDW state. (c) The normal and SDW state bands are compared.  Momentum positions for EDCs in Figs.~\ref{TD} and \ref{TD2} are marked on the top.}\label{compare}
\end{figure}

The electronic structure changes dramatically in the SDW state. Figure~\ref{SDW} shows the corresponding data of Fig.~\ref{normal} in the SDW state. There are now three hole-like Fermi surfaces near $\Gamma$, and  a cross-like spectral weight distribution near $M$  (Fig.~\ref{SDW}a). One more band, $\gamma'$, appears in the SDW state than in the normal state. Moreover, as will be highlighted  in Fig.\ref{compare} later, the energy positions of the bands  shift in comparison with the normal state. Figs.~\ref{SDW}d and \ref{SDW}e show the MDCs near $\Gamma$ and $M$ respectively, the bands clearly cross Fermi energy and do not show any sign of bending due to the opening of  any gap. We note the intensity of certain band may be weak at the Fermi level, but this is rather due to the matrix element effects, which has been the case for the normal state spectra in Fig.~1 already.

To understand the band structure in the SDW state, Figure~\ref{compare} plots band structures in the normal and SDW states together. As shown in Fig.~\ref{compare}b, the band folding mainly occurs  on the $\gamma$ band in the SDW state, causing the additional $\gamma'$ band. Both $\gamma$ and $\gamma'$ slightly push $\alpha$ and $\beta$ near $\Gamma$,  and they hybridize with $\delta$, generating a complicated band structure around $M$. Therefore they are slightly asymmetric with respect to the SDW zone boundary (the dash-dotted line). By comparing the bands in the normal and SDW states in Fig.~\ref{compare}c, pronounced band shifts could be identified. The   $\beta$ and $\delta$ bands shift downward, while the $\gamma$ band shifts upward near M. Such a large shift, especially the shift of a band well below Fermi surface, indicates that the band reconstruction is not due to any Fermi surface instability.

\begin{figure}[b!]
\includegraphics[width=8.5cm]{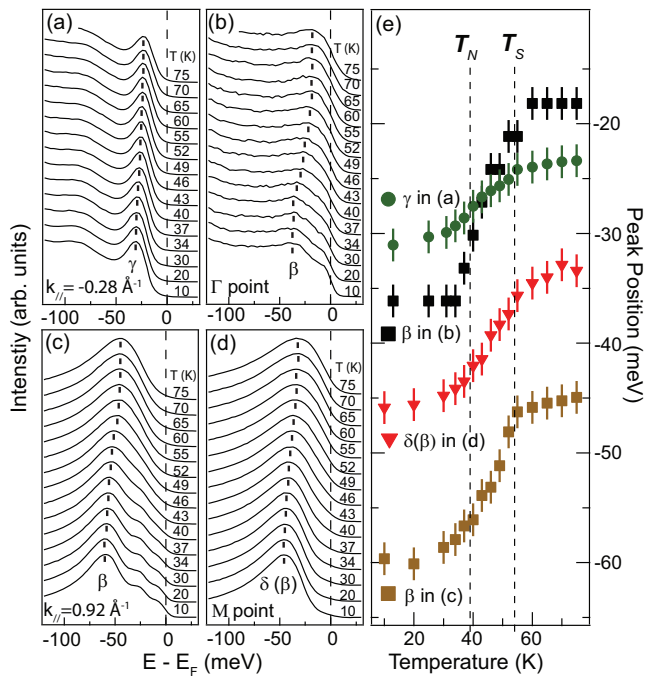}
\caption{(Color online) Temperature dependence of EDCs at selected momenta: (a) $k_x=-0.28$\AA$^{-1}$, (b) $k_x=0$ ($\Gamma$), (c) $k_x=0.92$\AA$^{-1}$, and (d) $k_x=1.1$\AA$^{-1}$ (M). (e) The summary of feature shifts in panels a-d.} \label{TD}
\end{figure}

To study the band reconstruction in detail, Figs.~\ref{TD}a-d show the temperature dependence of photoemission spectra at four representative momenta. It is clear that positions of various bands start to shift rapidly at the structural transition temperature $T_S$, and continue smoothly into the SDW state (Fig.~\ref{TD}e). For example, the $\beta$ band shifts downward by about 16~meV (Fig.~\ref{TD}c), in which a 12~meV  shift has already happened above the SDW transition temperature $T_N$, while half of the shift for the $\gamma$ band in Figs.~\ref{TD}a (since the folded $\gamma'$ is very weak) has happened above $T_N$.   The fact  that the bands shift smoothly across the SDW transition temperature can only be interpreted into two completely opposite ways:  either the magnetism is coupled to the electronic structure very weakly; or the electronic structure has been already strongly coupled to magnetic fluctuations above $T_{N}$.  However, the former case is unlikely. Since if such a large change in electronic structure were not related to the magnetic ordering, it must be caused by the lattice distortion or some parasitical secondary orbital/charge order. Neutron and x-ray scattering experiments have found that the lattice distortion in NaFeAs takes place at $T_S$, and saturates at $T_N$  with 0.36\% total lattice displacement. \cite{ShiliangLi,Parker}. It would cause about 1\% of change to the hopping parameters among various $d$ orbitals \cite{Harrison}. Since 1\% of the measured occupied bandwidth is just about 1~meV, the observed 12~meV shift between $T_{S}$ and $T_{N}$ is well beyond what the lattice distortion could cause.

\begin{figure}[t!]
\includegraphics[width=7cm]{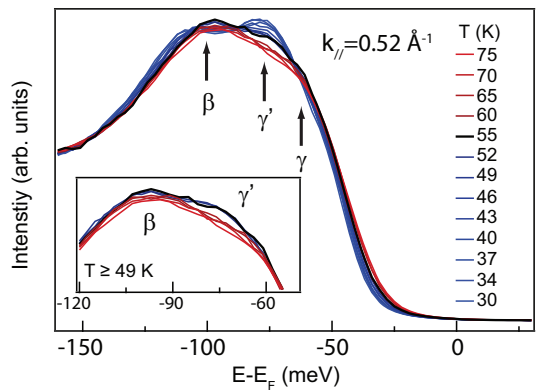}
\caption{(Color online) Temperature dependence of the EDC taken at     $k_x=0.52$\AA$^{-1}$. The inset is an enlargement of the top part of the spectra  taken near the structural transition temperature.} \label{TD2}
\end{figure}

The strong coupling between magnetism and electronic structure is further supported by the observation of band folding above the SDW transition in Fig.~\ref{TD2}. At $k_{\parallel}=0.52$\AA$^{-1}$, we expect that the $\gamma'$ band becomes visible in EDC when there is a band folding. Indeed, there is an abrupt change on the spectrum taken at 55~K; and the evolution below 55~K is rather smooth. Therefore, band folding occurs already at the structural transition temperature. Since the structural transition itself does not cause zone folding, this band folding indicates the existence of a short-ranged electronic ordering that locally doubles the unit cell, which can be captured by fast probes like ARPES. Because the evolution of band folding is smooth and no additional feature presents at the magnetic transition temperature,  we can conclude that  the short-ranged magnetic ordering must have developed  at $T_S$.    We note that  elastic neutron scattering would not detect such magnetic domains due to cancelation of the scattering from anti-phase domains. Furthermore, recent muon spin resonance ($\mu$SR) studies of NaFeAs could only detect SDW  below $T_N$  \cite{Parker}, indicating the fluctuation time scale of the magnetic domains are  faster than the $\mu$SR time scale above $T_{N}$.

The intimate relation between  the electronic structure reconstruction and the structural/magnetic orderings is  most remarkably represented by the fact that $T_{S}$ is not only the onset of lattice distortion, but also the onset of  the fluctuating SDW (as evidenced by the band folding) and electronic structure transition (the band shift). Our results clearly show that it is the electronic structure reconstruction that saves the total electronic energy significantly during both transitions. It is thus the driving force behind the magnetic transition and the rather soft structural transition (either through the SDW fluctuations or directly). In parallel, Fermi surface nesting can be ruled out as the driving force for the SDW in NaFeAs for two primary reasons. First, it is clear that the band reconstruction  goes beyond the nested (or semi-nested) portion of the Fermi surfaces near $\Gamma$ and $M$. In fact,  Fig.~\ref{compare} shows that most of the bands,  including the band far below the Fermi energy, participate in the reconstruction.  Second, no gap has been observed at the Fermi crossings.

The reconstruction of the band structure here shares many common aspects with that in the ``122" series of iron pnictides \cite{LXYang,YZhang,MYi}. The scale of the shift, ordered moment, structural transition amplitude, and ordering temperature  all  roughly scale with each other in these compounds. Therefore, the band reconstruction scenario discussed here is most likely universal for the structural and magnetic transitions  in all iron pnictides.

Models based on local magnetic exchange have successfully explained the subsequent transitions of lattice  and magnetism, and they suggest that the lattice transition is driven by magnetic fluctuations \cite{ChenFang,xu}.  Moreover, local exchange interactions have been suggested to  explain the observed band reconstruction, and they  do have similar energy scales  \cite{LXYang,YZhang,MYi}.
While our result can be viewed as a positive support of this picture,  it also emphasizes that the electronic structure has to be taken into account in order to fully understand the transitions.

To summarize, we have carried out a systematic photoemission investigation of high quality  NaFeAs single crystals, revealing the detailed electronic structure of the ``111" series of iron pnictides for the first time. We show that the band structure undergoes significant shifts, and short-ranged  magnetic ordering has already developed at the structural transition temperature. Our results suggest that electronic structure reconstruction, rather than Fermi surface nesting, plays a dominating role in causing both the lattice and magnetic transitions in iron pnictides, and support the existence of strong electron-electron correlation in these materials.

We acknowledge the helpful discussion with Dr. Dong-hui Lu. This work was supported by the NSFC, MOE, MOST (National Basic
Research Program No.2006CB921300), and STCSM of China.

\end{document}